\input{aipcheck}

\documentclass[
    ,final            % use final for the camera ready runs
  ]
  {aipproc}

\layoutstyle{8x11single}

\begin{document}

\title{Production and relevance of cosmogenic radionuclides in NaI(Tl) crystals}

% 29.40.Mc Scintillation detectors
% 95.35.+d  Dark matter (stellar, interstellar, galactic, and cosmological)
% 25.40.-h  Nucleon-induced reactions

\author{J. Amaré, S. Cebrián, C. Cuesta\footnote{Present address: Department of Physics, Center for Experimental Nuclear Physics and Astrophysics, University of Washington, Seattle, WA, USA}, E. García, C. Ginestra, M. Martínez\footnote{Present address: Università di Roma La Sapienza, Piazzale Aldo Moro 5, 00185 Roma, Italy}, M. A. Oliván, Y. Ortigoza, A. Ortiz de Solórzano, C. Pobes\footnote{Present address: Instituto de Ciencia de Materiales de Aragón, Universidad de Zaragoza-CSIC, Calle Pedro Cerbuna 12, 50009 Zaragoza, Spain}, J. Puimedón, M. L. Sarsa, J. A. Villar, and P. Villar}{
  address={Laboratorio de Física Nuclear y Astropartículas, Universidad de Zaragoza, Calle Pedro Cerbuna 12, 50009 Zaragoza, Spain\\
Laboratorio Subterráneo de Canfranc, Paseo de los Ayerbe s/n, 22880 Canfranc Estación, Huesca, Spain}
}

%\author{<author2>}{
%  address={<common address for author2 and author3>}
%}

%\author{<author3>}{
%  address={<common address for author2 and author3>}
%  ,altaddress={<author1 address>} % additional visiting address
%}

\begin{abstract}
The cosmogenic production of long-lived radioactive isotopes in
materials is an hazard for experiments demanding ultra-low
background conditions. Although NaI(Tl) scintillators have been used
in this context for a long time, very few activation data were
available. We present results from two 12.5 kg NaI(Tl) detectors,
developed within the ANAIS project and installed at the Canfranc
Underground Laboratory. The prompt data taking starting made
possible a reliable quantification of production of some I, Te and
Na isotopes with half-lives larger than ten days. Initial activities
underground were measured and then production rates at sea level
were estimated following the history of detectors; a comparison of
these rates with calculations using typical cosmic neutron flux at
sea level and a selected description of excitation functions was
also carried out. After including the contribution from the
identified cosmogenic products in the detector background model, we
found that the presence of $^{3}$H in the crystal bulk would help to
fit much better our background model and experimental data. We have
analyzed the cosmogenic production of $^{3}$H in NaI, and although
precise quantification has not been attempted, we can conclude that
it could imply a very relevant contribution to the total background
below 15 keV in NaI detectors.

\end{abstract}

\maketitle

%%%%%%%%%%%%%%%%%%%%%%%%%%%%%%%%%%%%%%%%%%%%
%% MAINMATTER
%%%%%%%%%%%%%%%%%%%%%%%%%%%%%%%%%%%%%%%%%%%%

\section{Introduction}

%% Intro ANAIS, set-up
The ANAIS (Annual Modulation with NaI(Tl) Scintillators) experiment
\cite{anais} aims at the confirmation of the DAMA/LIBRA annual
modulation positive signal, using the same target and technique at
the Canfranc Underground Laboratory (LSC). Following work on
previous prototypes \cite{bkg,analysis}, the ANAIS-25 set-up
\cite{anais25} consisted of two 12.5 kg NaI(Tl) crystals built by
Alpha Spectra; it was intended to understand and quantify the
background components and to assess the performance of the
detectors. Each module consists of a cylindrical NaI(Tl) crystal,
grown with selected ultrapure NaI powder and encapsulated in OFHC
copper. The coupling to Hamamatsu photomultiplier tubes was
performed in the clean room of LSC. ANAIS-25 was operated inside a
shielding consisting of 10 cm archaeological lead plus 20 cm low
activity lead.
%(named D0 and D1)

%measuremements
Detectors were built in Colorado and shipped to Spain; they arrived
to Canfranc on 27$^{th}$ November 2012 and were immediately stored
underground. Data taking started just three days afterwards,
allowing, together with the low background conditions and the very
good energy resolution of the detectors, a precise study of the
isotopes induced in the crystals by the exposure to cosmic rays;
details of this work can be found at \cite{jcap} and are summarized
here. Data analyzed investigating the cosmogenic activation
correspond to different data sets. Data sets I and II, acquired in
slightly different gain conditions, span altogether for 210 days
from the beginning of the data taking to the end of June 2013. Set
III includes 88 days of data from March to June 2014, once the
contribution of most of the cosmogenic nuclei was significantly
reduced, allowing to identify longer-living products. Possible
presence of tritium was evidenced in data taken from June 2014 to
March 2015 and will be also discussed.

\section{Iodine, Tellurium and Sodium isotopes}

\begin{figure}
  \includegraphics[height=.25\textheight]{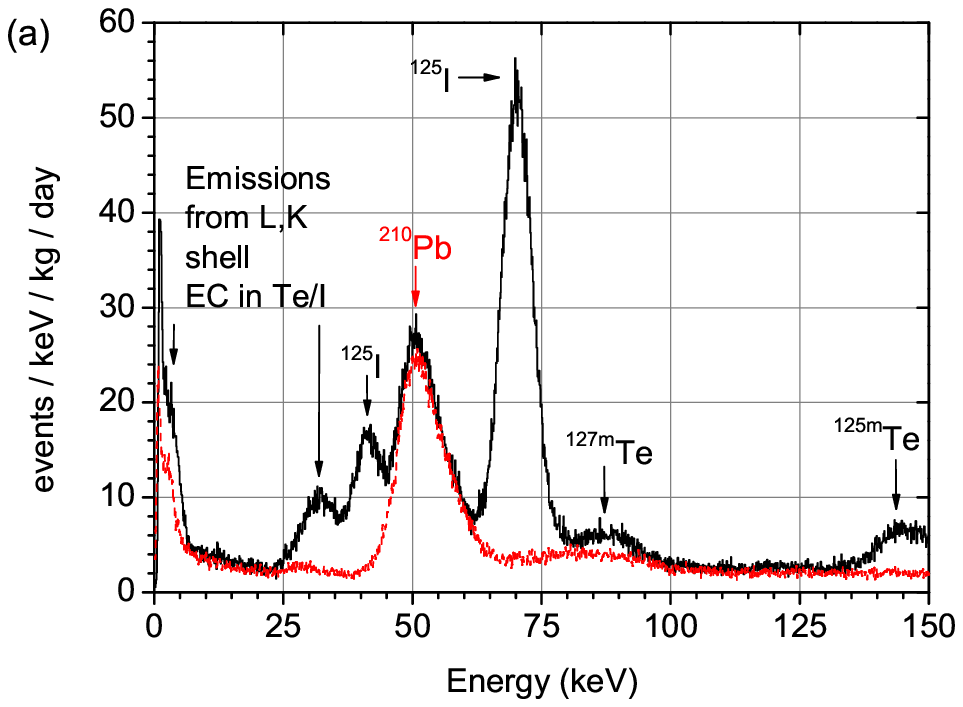}
  \includegraphics[height=.25\textheight]{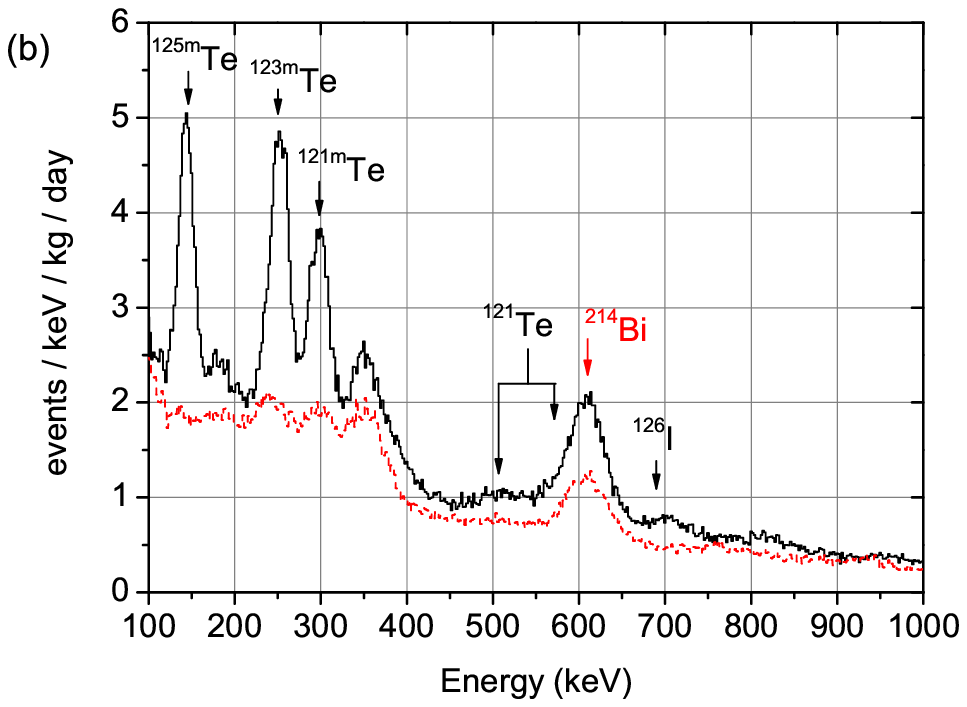}
  \caption{Comparison of the ANAIS-25 spectra evaluated in the beginning of data set I (black solid line) and about fifteen months afterwards, during data set III (red dashed line). The low (a) and high (b) energy regions are shown. Several cosmogenic emissions have been identified and are labeled (in black) together with the main background lines (in red).}
  \label{spectra}
\end{figure}

Figure~\ref{spectra} compares the spectra of one of the ANAIS-25
detectors evaluated in the beginning of data set I and about fifteen
months afterwards, during data set III, over 13.0 and 18.5 days,
respectively. Several lines can be clearly attributed to cosmogenic
activation and signatures from $^{125}$I, $^{127m}$Te, $^{125m}$Te,
$^{123m}$Te and $^{121m}$Te were directly identified in energy
spectra. $^{126}$I and  $^{121}$Te signals were evaluated by
analyzing coincidences between detectors. Although not directly seen
in figure~\ref{spectra}, $^{22}$Na, having a longer mean life than I
and Te products, was also identified and quantified using
coincidence spectra along data set III. The cosmogenically induced
activity, which is the initial activity A$_{0}$ corresponding to the
moment of storing crystals deep underground at LSC, was deduced
studying the exponential decay of the identifying signature produced
by each I and Te isotope considering data sets I and II. For
$^{121}$Te, production from $^{121m}$Te decay was properly taken
into account. Concerning $^{22}$Na, the integrated signal along data
set III was used. Production rates R$_{p}$ at sea level were
estimated from measured A$_{0}$ values, considering that saturation
was reached at detector production in Colorado (where the cosmic
neutron flux is 3.6 times higher than at sea level) and a further
exposition of 30 days for the boat trip from US to Spain
\cite{jcap}. Table~\ref{A0rates} summarizes the results obtained for
A$_{0}$ and R$_{p}$ for all the identified cosmogenic products.

Production rates were also calculated considering the cosmic neutron
spectrum at sea level from \cite{gordon} and selecting carefully the
excitation functions to minimize deviation factors between available
measurements and results from different libraries and calculations
from semiempirical formulae \cite{jcap}. Table~\ref{A0rates} shows
also these computed production rates. Remarkable agreement with
experimental rates has been achieved, specially for the products
having excitation functions well validated against measurements.

\begin{table}
\begin{tabular}{lcccccc}
\hline Isotope & $T_{1/2}$  &  $A_{0}$  & Excitation function &
Cal R$_{p}$ &  Exp R$_{p}$ & Cal$/$Exp \\
& (days)&  (kg$^{-1}$d$^{-1}$) & & (kg$^{-1}$d$^{-1}$) & (kg$^{-1}$d$^{-1}$) & \\
\hline

$^{126}$I & 12.93$\pm$0.05  &  430$\pm$37 & MENDL-2+YIELDX   & 297.0 & 283$\pm$36  & 1.1 \\
$^{125}$I  & 59.407$\pm$0.009 &  621.8$\pm$1.6 & TENDL-2013+HEAD-2009 &  242.3  & 220$\pm$10 & 1.1  \\
$^{127m}$Te  & 107$\pm$4 & 32.1$\pm$0.8 &TENDL-2013+extrapolation   &  7.1 &  10.2$\pm$0.4 & 0.7 \\
$^{125m}$Te  & 57.40$\pm$0.15 &  79.1$\pm$0.8 &TENDL-2013+HEAD-2009   &  41.9 &  28.2$\pm$1.3 & 1.5 \\
$^{123m}$Te   & 119.3$\pm$0.1 & 100.8$\pm$0.8 & TENDL-2013+HEAD-2009 & 33.2  &  31.6$\pm$1.1 & 1.1 \\
$^{121m}$Te  & 154$\pm$7 & 76.9$\pm$0.8 & TENDL-2013+HEAD-2009 & 23.8  &  23.5$\pm$0.8 &1.0 \\
$^{121}$Te & 19.16$\pm$0.05 & 110$\pm$12       & TENDL-2013+YIELDX  &  8.4  & 9.9$\pm$3.7 & 0.8 \\
$^{22}$Na & (2.6029$\pm$0.0008) y   & 159.7$\pm$4.9 & TENDL-2013+YIELDX &  53.6 & 45.1$\pm$1.9 &  1.2\\

 \hline
\end{tabular}
\caption{Results derived for each cosmogenic product: initial
activities underground $A_{0}$, production rates R$_{p}$ obtained
experimentally and calculated using different selections of the
excitation function. Last column presents the ratio between the
calculated and the experimental rates.} \label{A0rates}
\end{table}

\section{The case of tritium}

\begin{figure}
  \includegraphics[height=.26\textheight]{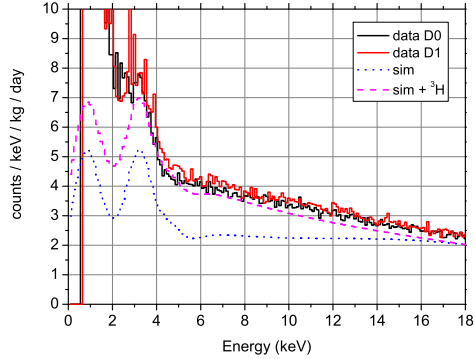}
  \caption{The very low energy region of the energy spectrum measured in ANAIS-25 detectors (solid lines) compared with the simulated one considering all the known and quantified intrinsic and cosmogenic activities in the detectors and main components of the set-up (dotted line) and adding also an arbitrary activity of $^{3}$H in the NaI crystal (dashed line).}
  \label{tritium}
\end{figure}

The contribution of the cosmogenic products to the background of the
ANAIS-25 detectors was evaluated by Geant4 simulation, considering
the measured activity values A$_{0}$ \cite{jcap}. Only $^{22}$Na,
producing a peak around 0.9~keV, is relevant for dark matter
searches. This contribution was included in the overall background
model of the detectors, together with that of the known activities
of the crystal and other components of the set-up (see details at
\cite{clara}). When the simulated spectrum including all well-known
contributions is compared with the one measured, a very good
agreement was obtained except for the very low energy region. But
this problem can be solved if an activity of $\sim$0.2 mBq/kg of
$^{3}$H homogeneously distributed in the NaI crystal is added to the
model, as shown in figure~\ref{tritium}.

\begin{figure}
  \includegraphics[height=.25\textheight]{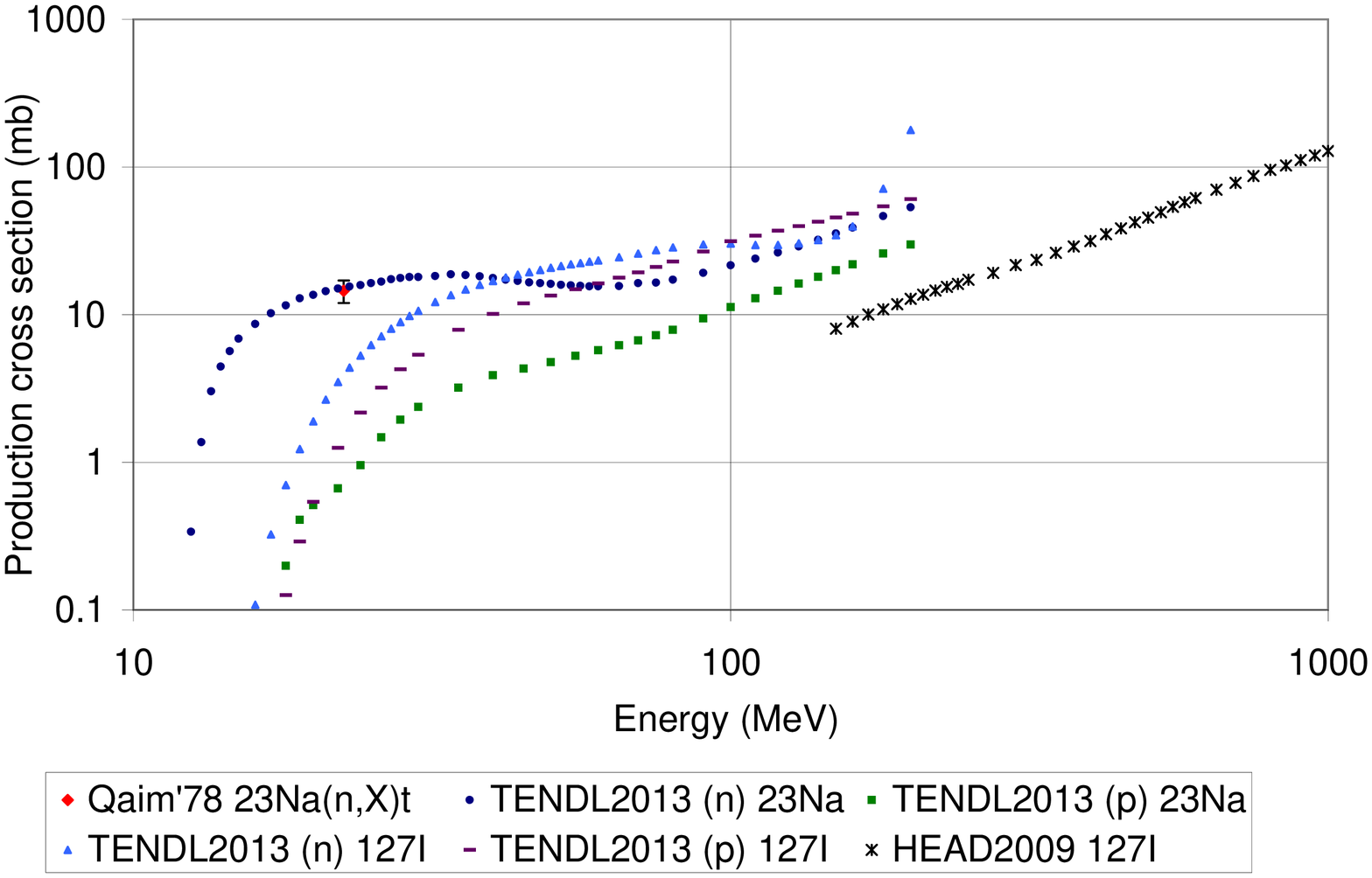}
  \caption{Comparison of excitation functions for production of $^{3}$H on $^{23}$Na and $^{127}$I by nucleons (neutrons are indicated as n and protons as p) taken from different sources
(see text).}
  \label{eftritium}
\end{figure}

This required activity is about twice the upper limit set for
DAMA/LIBRA crystals \cite{damalibra} and lower than the saturation
activity predicted assuming the production rate of $^{3}$H in
NaI(Tl) of 31.1~kg$^{-1}$d$^{-1}$ calculated in \cite{mei} using
production cross section estimated with TALYS 1.0 code. An attempt
has been made to quantify this production rate in $^{23}$Na and
$^{127}$I (both having 100\% natural isotopic abundance in NaI)
using the same approach followed for the cosmogenic isotopes
identified in ANAIS-25. First, available information on the
excitation function by nucleons was collected, as shown in
figure~\ref{eftritium}: only one experimental result was found in
the EXFOR database \cite{exfor} and cross sections were taken from
the TENDL-2013 (TALYS-based Evaluated Nuclear Data Library) library
\cite{tendl} up to 200 MeV and from the HEAD-2009 (High Energy
Activation Data) library \cite{head} from 150 to 1000 MeV. Then, the
production rate was computed convoluting a selected excitation
function with the energy spectrum of cosmic neutrons at sea level,
using the parametrization from \cite{gordon}.
Table~\ref{ratestritium} summarizes the rates for several
descriptions of the cross sections. The total rate considering data
from TENDL-2013 library below 150 MeV reproduces the value obtained
in \cite{mei}, since the library is also based on TALYS code.
Although information on the excitation function is very limited, it
seems that the contribution to the production rate from energies
above 150 MeV is not negligible. Assuming that in this high energy
range neutron and proton cross sections are comparable and that
production from $^{23}$Na and $^{127}$I is similar (as for energies
below 150 MeV, see table~\ref{ratestritium}) the production rate
could be of $\sim$50 kg$^{-1}$d$^{-1}$ summing the contributions in
table~\ref{ratestritium}. For such a rate, an exposure of 1.9 y to
the neutron flux at Grand Junction, Colorado, would produce the
required $^{3}$H activity in ANAIS-25 crystals.

\begin{table}
\begin{tabular}{lccc}
\hline
Isotope  & Library & Energy range & Production rate
  (kg$^{-1}$d$^{-1}$)  \\
\hline
$^{23}$Na & TENDL-2013 \cite{tendl} & $<$150 MeV & 14.3 \\
$^{127}$I & TENDL-2013 \cite{tendl}& $<$150 MeV & 15.4 \\
$^{127}$I & HEAD-2009 \cite{head} & 150-1000 MeV & 10.9 \\
\hline
\end{tabular}
\caption{Production rates of $^{3}$H in NaI isotopes obtained
considering different excitation functions.} \label{ratestritium}
\end{table}

\section{Summary}

Cosmogenic activation of NaI(Tl) crystals was evaluated from
measurements taken at the Canfranc Underground Laboratory using two
12.5 kg detectors produced by Alpha Spectra in Colorado (US) and
shipped to Spain \cite{jcap}. Production rates at sea level of a few
tens of nuclei per kg and day for Te isotopes and $^{22}$Na and of a
few hundreds for I isotopes were derived from measured activities,
in good agreement with calculations. In addition, presence of
$^{3}$H seems plausible. I and Te isotopes are not relevant after
short cooling underground, but $^{22}$Na and $^{3}$H, due to their
longer mean lives and low energy emissions, affect the region of
interest for a dark matter search. For ANAIS-25 detectors, their
contribution from 1 to 10 keV is estimated to be of 1.4 and 13.2
events$/$kg$/$d, respectively.

%%%%%%%%%%%%%%%%%%%%%%%%%%%%%%%%%%%%%%%%%%%%%%%%
%% BACKMATTER
%%%%%%%%%%%%%%%%%%%%%%%%%%%%%%%%%%%%%%%%%%%%%%%%

\begin{theacknowledgments}
This work was supported by the Spanish Ministerio de Economía y
Competitividad and the European Regional Development Fund
(MINECO-FEDER) (FPA2011-23749), the Consolider-Ingenio 2010
Programme under grants MULTIDARK CSD2009-00064 and CPAN
CSD2007-00042, and the Gobierno de Aragón (Group in Nuclear and
Astroparticle Physics, ARAID Foundation and C.~Cuesta predoctoral
grant). C.~Ginestra and P.~Villar are supported by the MINECO
Subprograma de Formación de Personal Investigador. We also
acknowledge LSC and GIFNA staff for their support.
\end{theacknowledgments}

\bibliographystyle{aipproc}

\begin{thebibliography}{9}

\bibitem{anais}
J.~Amaré et al., \emph{Physics Procedia} \textbf{61}, 157--162 (2015).
%From ANAIS-25 towards ANAIS-250

\bibitem{bkg}
S.~Cebrián et al., \emph{Astropart. Phys.} \textbf{37}, 60--69
(2012).
%Background model for a NaI (Tl) detector devoted to dark matter searches,

\bibitem{analysis}
C.~Cuesta al.,  \emph{Eur. Phys. J. C} \textbf{74}, 3150 (2014).
%Bulk NaI(Tl) scintillation low energy events selection with the ANAIS-0 module,

\bibitem{anais25}
J.~Amaré et al., \emph{Nucl. Instrum. Meth. A} \textbf{742},
187--190 (2014).

\bibitem{jcap}
J.~Amaré et al., \emph{JCAP} \textbf{02}, 046 (2015).
%Cosmogenic radionuclide production in NaI(Tl) crystals

\bibitem{gordon}
M.~S.~Gordon et al, \emph{IEEE Trans. Nucl. Sci.} \textbf{51},
3427--3434 (2004).

\bibitem{clara}
C.~Cuesta et al, ``Background analysis and status of the ANAIS dark
matter project'', in these proceedings.

\bibitem{damalibra} R.~Bernabei et al., \emph{Nucl. Instrum. Meth. A} \textbf{592}, 297
(2008).

\bibitem{mei}
D.~M.~Mei et al, \emph{Astropart. Phys.} \textbf{31}, 417--420
(2009).

\bibitem{exfor}
Experimental Nuclear Reaction Data (EXFOR)
\url{http://www.nndc.bnl.gov/exfor/exfor.htm},
\url{http://www-nds.iaea.org/exfor/exfor.htm}.

\bibitem{tendl}
A.~J.~Koning and D.~Rochman, \emph{Nuclear Data Sheets}
\textbf{113}, 2841 (2012),
\url{http://www.talys.eu/tendl-2013.html]}
%Modern Nuclear Data Evaluation With The TALYS Code System

\bibitem{head}
Y.~A.~Korovin et al, \emph{Nucl. Instrum. Meth. A} \textbf{624},
20--26 (2010).
% High Energy Activation Data Library (HEAD-2009)


%\bibitem{BrownAustin:2000}
%M.~P. Brown,  and K.~Austin, \emph{Appl. Phys. Letters} \textbf{85},
%  2503--2504 (2000).

%\bibitem{Wang}
%R.~Wang, ``Title of Chapter,'' in \emph{Classic Physiques}, edited by
%  R.~B. Hamil, Publisher Name, Publisher City, 2000, pp. 212--213.

%\bibitem{SJ:1999}
%C.~D.~Smith and E.~F.~Jones,  ``Load-Cycling in Cubic Press,'' in
%  \emph{Shock Compression of Condensed Matter-1999}, edited by M.~D.~F. et~al.,
%  AIP Conference Proceedings 505, American Institute of Physics, New York,
%  1999, pp. 651--654.

\end{thebibliography}

\end{document}